\documentclass{vgtc}
\pdfoutput=1
\usepackage{algorithm}
\usepackage{algorithmic}
\usepackage{mathptmx}
\usepackage{graphicx}
\usepackage{times}
\usepackage{url}
\usepackage{subfigure}
\usepackage{fancyvrb}
\usepackage{multirow}
\usepackage{hhline}
\usepackage[labelsep=space]{caption}

\newcommand{\Tool}{\textsc{ParaDTI}}

\title{Depth-dependent Parallel Visualization with 3D Stylized Dense Tubes}

\author{Haipeng Cai, Jian Chen and Alexander P. Auchus}
\authorfooter{
\item
 Haipeng Cai is with University of Southern Mississippi, e-mail: haipeng.cai@eagles.usm.edu.
\item
 Jian Chen is with University of Southern Mississippi, e-mail: jian.chen@usm.edu.
\item
 Alexander P. Auchus is with University of Mississippi Medical Center, e-mail: aauchus@umc.edu.
}

\shortauthortitle{Haipeng Cai\MakeLowercase{\textit{et al.}}: {\Tool}}

\abstract{
We present a parallel visualization algorithm for the illustrative rendering of depth-dependent stylized dense tube data
at interactive frame rates. While this computation could be efficiently performed on a GPU device, we target a parallel framework to enable it to be efficiently running on an ordinary multi-core CPU platform which is much more available than GPUs for common users. Our approach is to map the depth information in each tube onto each of the visual dimensions of shape,
color, texture, value, and size on the basis of Bertin's semiology theory. The purpose is to enable more legible
displays in the dense tube environments. A major contribution of our work is an efficient and effective parallel
depth-ordering algorithm that makes use of the message passing interface (MPI) with VTK. We evaluated our framework with
visualizations of depth-stylized tubes derived from 3D diffusion tensor MRI data by comparing its efficiency with several other alternative parallelization platforms running the same computations. As our results show, the parallelization framework we proposed can efficiently render highly dense 3D data sets like the tube data and thus is useful as a complement to parallel visualization environments that rely on GPUs.
}


\CCScatlist{
\CCScat{Computer Graphics}{I.3.1}{Parallel processing}
\CCScat{Computer Graphics}{I.3.4}{Graphics Utilities}
}

\renewcommand{\manuscriptnotetxt}{}

\begin{document}
\setlength{\pdfpagewidth}{8.5in} 
\setlength{\pdfpageheight}{11in}

\firstsection{Introduction}
\label{sec:intro}

\maketitle
\pagenumbering{arabic}
  \pagestyle{plain}
  \thispagestyle{plain}

When visualizing large-scale geometrical data such as dense tubes, one of the critical issues is the visual perception
in the depth dimension due to inherent clutters or occlusions as the result of overlapping graphical signs or
structures. In order to improve the overall visual legibility from the prospective of depth perception for
three-dimension data, mapping depth information to various visual variables of graphical representations can be
effective means for enhancing depth perceptions in the dense data visualizations.

On the basis of Bertin's semiology
theory ~\cite{Ber83}, the primary visual variables to be mapped could include size, color, value and transparency, etc. For instance, a linear mapping from per-vertex depth value to the tube radius (i.e. the size) in the visualization of dense 3D tubes will give viewers a visual cue for discerning depth positions when the radii are gradually decreasing along the viewing
direction. Similarly, a consistent mapping from depth to color provides a constant correspondence between distance of geometry and
color value thus helps viewers to orient along the depth dimension. In either case, better depth perception is conducive
to improvement in overall legibility of the visualized data.

To make the depth-dependent visualizations interactive, real-time computation involved in the depth mappings is
required. There are two essential compute-intensive steps to be executed every time the depth reordering is needed, for
example when the data view is rotated. First, depth values are calculated according to the updated viewing direction
and then sorted along that direction. Second, mappings are computed and then the data are rendered over again to update
the visualization. Concisely, depth sorting and re-rendering should be performed once depth order is shuffled as typical
result of data transformation that changes the order. In order to obtain an interactive frame rate of such
visualizations, therefore, these computations are required to finish in real-time, which have been proven difficult to achieve by our tests, however, with either sequential method or by direct use of currently available facilities such as VTK with its parallelism support.

With this performance challenge, it is reasonable to consider parallelization of the depth-dependent visualization
described above to make it interactive. While GPUs are being increasingly applied in many modern parallel computations, and indeed, visualizations of large-scale dense geometry data could be a perfect fit for GPU computing platforms, we aim at a cheaper solution to the same challenges. In particular, we target a solution that can be a useful complement to the GPU computing paradigm when the GPU devices and related high-end hardware configurations are not readily available.
In fact, this is mostly true since GPUs are generally much more expensive than ordinary computer users would like to afford for the tasks like the dense tube data visualizations we discuss in this paper, which can indeed be completed with cheap PC hardware using our parallelization framework.

This paper describes a parallel visualization method that supports real-time
computations for interactive depth mappings by using the message passing interface (MPI) with VTK while extending
current VTK facilities for the purpose of performance optimization. Through the optimized coordination between a parallel
depth ordering algorithm and parallel rendering method plus customized data structures for real-time depth mappings, our
approach has been verified to be efficient in the visualization scenarios we described by its application to 3D dense tube
data.

We have applied the method proposed to depth-dependent 3D dense tube visualizations with depth mappings to all the primary visual
variables mentioned before and have obtained interactive rendering speed with either single variable mapping applied or
multiple variable mappings freely combined. It is noticeable that even with the combination of mappings from depth to
size and those to any other visual variables, in which two passes of depth sorting and rendering plus tube generation from
polylines are all required for each frame, our approach has still been able to render the dense data sets at interactive frame rates.

\section{Related Work}
In this section, we describe previous work most relevant to our parallel visualization method. Related past work can be
classified into two categories: depth enhancement and parallel visualization.
\subsection{Depth Enhancement}
There have been many previous work dedicating to technical solutions to the depth perception issues and visual
occlusions in 3D data visualizations. To name a few, a rich set of landmarks and context cues ~\cite{LFH06} and
shading and transparency ~\cite{II02} both contribute in enhancing visual perception in the depth dimension while alleviating
occlusion problems within overlapping structures. Focusing on strengthening depth perception, Bruckner et al. employ
volumetric halos to improve the 3D legibility of visualized volume data ~\cite{BG07}. They introduce
different halos according to different ways of halo-volume combination and use halos to construct inconsistent lighting,
which accentuates depth even further from another aspect.

Elmqvist et al. ~\cite{ET08} give an ever complete discussion about occlusion management in 3D visualization where they focused on reducing 3D occlusions. Occlusion management for visualization is a more general class of visibility problem in computer graphics, which is concerned with improving human perception for specialized visual tasks such as occlusion, size and shape. This method extensively helped improve the legibility of 3D data visualizations. In contrast, we investigate how to manipulate typical retinal variables in graphics perception to help
achieve a better depth legibility.

Even direct volume rendering techniques often suffer from poor depth cues because the data sets commonly have a large number of overlapping structures. With MIP (maximum intensity projection) rendering~\cite{DV10}, however, only few effort is required to create a good understanding of the structures represented by high signal intensities. This algorithm adds two different visual cues, occlusion revealing and depth based color. In the first one, they modify the MIP color in the presence of occluding objects with the same materials than the one at the point of maximum intensity while in other the actual position of the shaded fragment is used to change its color using a supporting spherical map. In this paper, we explore depth enhancement in dense geometry visualizations by encoding depth information with various visual variables.

Ritter et al.~\cite{RCV06} employ hatching strokes to communicate shape while using distance-encoded shadow to further enhance depth perception in their vascular structure visualization. In addition, they achieve a real-time performance using GPU-based
hatching algorithm, which is efficient for rendering complex tabular structures with depth being emphasized.
Similarly, we handle tabular shapes in our visualization scenario but intend to improve depth perception in a much dense
3D tube geometries derived from human brain MRI data. Also, we we are to provide a cheaper interactive rendering
solution on common multi-core CPU than the GPU rendering they have employed.

\subsection{Parallel Visualization}
Parallelization has been extensively harnessed in visualization scenarios where performance becomes a challenge. In
~\cite{JCWKM00}, the authors developed a scalable and portable parallel visualization system based on augmenting VTK for efficiently visualizing large scale time-varying data. The system they proposed provides parallelism on both task and pipeline level and primary addressed to visualization programmers. Also at a system scale but even earlier, SCIRun ~\cite{JP99} had
offered task and data parallelism as a data flow based visualization system running on shared-memory machine with
multiprocessors. This system was extended to support task parallelism on distributed-memory architectures ~\cite{MHJ98}.
We present a light-weighted parallelization method for large geometry visualization by using existing facilities like
MPI and VTK instead of providing a fully featured system or extended programming library.

Compared to the system level solution, a lot more parallelization efforts for visualization focus on parallel rendering, ranging from photo-realistic rendering ~\cite{RCJ98}, volume rendering ~\cite{Wit98} to parallel iso-surfacing ~\cite{PSLHS98}.
Among a large set of previous work specific to parallel polygon rendering, Crockett ~\cite{CO94} harnessed message
passing architectures for polygon rendering parallelism that reduces memory usage and network contention while
overlapping computation and communication. He also gives an overview of parallel rendering techniques from both hardware
and software prospectives later on ~\cite{Cro97}.

Other researchers have probed different indirect aspects, such as image composition schemes ~\cite{LRN96} and data decomposition strategy ~\cite{Whi94}, to improve polygon rendering performance. More recently, various parallel rendering algorithms including sort-first, sort-last and the hybrid of them, were evaluated when being used on shared-memory computers while these algorithms are originally targeting distributed-memory architectures ~\cite{NAWS11}.
In our work, we explore polygon rendering parallelization and employ image compositing as well but
particularly serve depth enhancing thus provide a more legible 3D geometry visualization by overlapping parallel depth
sorting and parallel polygonal data rendering.

Note that the parallel sorting problem~\cite{leiserson2001introduction,donald1999art}, which is at the core of our parallelization framework here, would be easily solved in highly efficient on GPU computing platforms with extensive existing algorithms available~\cite{pharr2005gpu,sintorn2008fast}. In this paper instead, we target a cheaper solution without relying on high-end computing resources such as GPUs. Alternatively and complementarily, we use CPU-based parallel sorting algorithms leveraging a single processor of multiple cores, which has been almost a bottom-line configuration for modern common computers.

\section{Our Methods}
\subsection{Depth Dimension Management}
According to Bertin's theory, visual legibility of two-dimensional (2D) graphical representations can be characterized by graphical density, angular separation and retinal separation ~\cite{Ber83}. Further, retinal separation is defined by six visual
variables including size, color, shape, value, orientation and texture. Illuminated by the Bertin's legibility rules
described in terms of these dimensions, it is promising to explore the visual legibility issues in 3D data
visualizations by examining 3D legibility dimensions. Although, our exploration is still based on the legibility
framework proposed by Bertin, certain expansion is required to fully characterize legibility in 3D graphics representations.

While Bertin's legibility dimensions serve 2D graphical representations, there is a lack of such
dimensions for legible 3D data visualizations. We expanded Bertin's framework from 2D to 3D by adding depth separation
dimension that is characteristic of 3D data and examine how typical retinal variables effect legibility of 3D
visualizations by investigating visual encodings that map depth information to each one of those variables examined respectively.
By such encodings, users are given visual cues to better discern depth locations thus the overall legibility of
visualization can be enhanced. Among others, currently we study three variables, size, color and value, which are inherited
from Bertin's framework directly and an extended variable, transparency, which is an important factor influencing depth
perception in 3D geometry. In our visual encodings, depth information of geometries can be either encoded by a single
visual variable alone or by multiple variables combined. By comparing different encodings, it would be revealed how
those visual variables effect the depth separation dimension hence the overall 3D data legibility.

\subsection{The Parallel Visualization Pipeline}
\begin{figure*}[htb]
  \centering
  \includegraphics[width=1.0\linewidth]{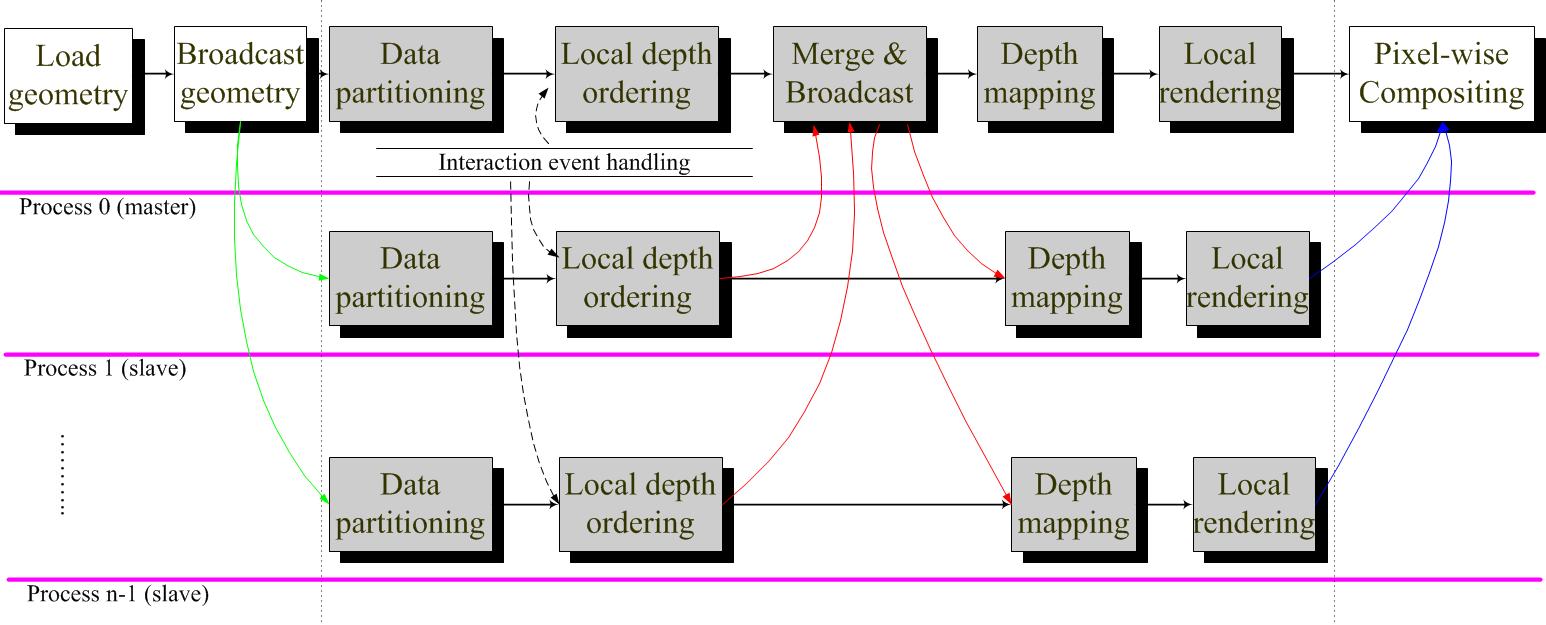}
  \caption{\label{fig:pipelineOutlook}
  The overview of our parallel visualization pipeline.}
\end{figure*}

Our parallel visualization pipeline is outlined in Figure~\ref{fig:pipelineOutlook}. The parallelism is powered by MPI and
visualization by VTK with parallelization support. Among the four processes, the master process $P_0$ is responsible for
data I/O, visualization interactions and coordinations required for a parallel rendering with consistent depth mappings
besides for rendering local data partitions as all other slave processes. The collaborations between the master
process and slave processes involve all key steps in the pipeline from data decomposition to parallel depth sorting
and geometry rendering.

\subsection{Data Decomposition}
Data decomposition is ordinarily an essential part for a parallelization mechanism. Although the concrete decomposition
scheme can be very much dependent of the interrelations between data components and there are different levels of
granularity of the data components as they are defined, it makes sense to split the whole data set into independent
partitions as such that data processing of each partition can be performed in parallel. In the case of 3D tubes, for
instance, a single tube is regarded as the minimal component and vertices on a tube will not be assigned to different partitions.

In addition, to maximally harness the computing resources available, we decompose the whole geometry model in simply an
average manner and then evenly distribute the computational tasks for sorting, mapping and rendering to all processes.
This simple data partitioning is efficient for our tube case because for one thing there is no data or semantic dependency
among all tubes, and for another, task load for each process is closely equal to others even if the master process will be
assigned certain managing roles. However, general data decomposition itself is a separate topic and there is no
generally optimal solutions, which are out of the scope of this paper.

When using MPI as the underlying parallel run-time support, the above data are decomposed according to the local process
id ($LocalProcId$) and total number of processes specified ($ProcNum$). Precisely, given all the data components ${C_0,
C_1,C_2,\cdots,C_{n-1}}$ in the equally-partitioning scheme, local sub-range of data for process $i$ will be
${C_{sidx},C_{eidx}}$ where $sidx=n/ProcNum*LocalProcId, eidx=n/ProcNum*(LocalProcId+1)$. Specially, the last process
may take more or less data components than others if $n$ is not exactly divided by $ProcNum$ when $eidx=n$.
Figure~\ref{fig:partitioning} illustrates this data decomposition scheme while showing the overall picture of the
depth-stylized visualization is rendered in parallel.

\subsection{Parallel Depth Sorting}
\subsubsection{Per-vertex Depth Ordering} 
In application scenarios like our 3D stylized dense tube visualization, mappings from depth information of each
vertex (or other unit of geometry like triangles or stripes) to visual attributes, such as size, color and transparency
that are referred as retinal variables in Bertin's semiology theory, should be consistent regardless of the current user
viewing directions in order to serve depth perception hence visual legibility along the depth dimension of the visualization.
In other words, all the vertices (or other geometry units) need to be ordered along the current viewing direction in
order to be consistently mapped to visual attribute values. That being said, once the viewing direction changes,
typically occurring when users rotate the data view, those vertices need to be reordered before mappings take place over
again to refresh the visualization.

In our dense tube environment, vertex-wise depth ordering is required for a
depth-dependent tube size assignment with which a better depth perception of even a single 3D tube can be obtained when, for
instance, a tube tapers or grows in its radius along the depth direction.

\subsubsection{Real Time Sorting} 
According to the necessity of depth ordering explained above, real-time depth mapping relies on real-time depth sorting.
For the per-vertex depth sorting, this computation is essentially the sorting of a sequence of floating-point numbers. For
depth sorting of other geometry units, the depth sorting is often eventually reduced to a per-vertex depth sorting
problem as well. For instance, if we only want to discern the depth locations at the level of tube rather than that of
vertex (thus the visual variable value of vertices on a tube is always the same), a vertex can be selected to
represent the tube and then the per-tube sorting is eventually reduced to the depth ordering of the representative
vertices.

Therefore, we generalize the depth sorting problem into the sorting of a sequence of cells. Practically, the
cell can either be a single value such as floating-point number or a packed data such as a struct including multiple
fields. We use the cell array for depth mapping in this paper.

While there are a rich set of parallel sorting algorithms freely available~\cite{ParallelSortUrl}, they generally
serve the solitary purpose of sorting and are usually implemented as stand-alone parallel applications. Since our ultimate
goal is to parallelize dense geometry visualization in which depth mappings are integrated, we need a holistic parallel
framework in which the sorting algorithm works together with other steps such as depth mapping and parallel rendering in
such a way that the overall visualization performance can be maximized. Our parallel sorting algorithm has been
accommodated to parallel rendering (see section~\ref{sec:ParallelGeometryRendering}) for which data partitioning is involved,
and optimized to mesh with efficient depth mappings (see section~\ref{sec:DepthMappings}).

We adopt mixed sorting algorithm for our parallel depth sorting by the following key steps. Firstly, each process
updates the depth values (z-coordinates) of local vertices through a simple vector arithmetic with the current camera parameters (focal point and position, etc). Then, every single process sorts vertex depth values in the partition assigned using a typical
quick sort algorithm and sends the sorted depth information to master process once finished. Finally, the master process
gathers locally sorted partitions and performs either a multi-way merge sort or multiple two-way merge sort. We employed the latter
merge sort scheme on the master process since it is more efficient as an iterative two-way merging can be performed once a
sorted partition is received from a slave process without waiting all processes to finish local sorting.

Algorithm~\ref{sortingmapping} shows how this parallel sorting algorithm works while illustrating the real-time depth
mappings fit for the parallel visualization framework as a whole.

\subsection{Parallel Geometry Rendering}
\label{sec:ParallelGeometryRendering}
In the application scenarios like our stylized tube visualization, the primary performance challenges come from two
sources, namely depth sorting and geometry rendering. For each updated frame, the whole geometry model needs be rendered
over again after depth sorting to reflect the depth mapping updates. Although both are critical for a real-time update,
the depth sorting time ($T_s$), compared to the rendering time ($T_r$), takes only a minor proportion of the whole frame
update time ($T=T_s+T_r$).

According to our sample test with a geometry of 140,000 vertices, $T_s/T$ is strictly less than
$0.1$ in the per-vertex depth-colored visualization. This shows that the rendering part is a bottleneck for
the overall visualization performance. In other word, interactive depth-stylized visualization depends on real-time rendering
of the depth-mapped geometries. Instead of involving GPU computation that is dependent of graphics hardware
architecture, We explore the parallelism in the dense geometry rendering by harnessing multiple computing hardware resource
thus provide a cheap yet more readily applicable parallel rendering solution.

Our approach to rendering in parallel is simply consisted of two main steps. After data decomposition, we firstly deploy
the local partition to each of the $n$ processes. Here a process is a general computing unit that can either be a
single processor on multiple-processor platform, a single core on a multi-core processor or a worker thread on a single
core processor. For each rendering frame, all separate renditions with each done by a single process are aggregated into
a single complete rendition that is only visible on the one of the processes randomly elected as the master process. This
aggregation is practically conducted by means of pixel-wise image compositing as the second step detailed as follows.

\subsubsection{Pixel-wise Compositing}
when each process finished the local rendering of the partition assigned (partial geometry), the rendition is
essentially a set of pixels in the frame buffer. As such, pixel-wise compositing is actually a process of compositing
frame buffers. In practice, to reduce computational costs, compositing only the color buffer and depth buffer is
sufficient for our visualization purpose. For simplicity, we describe compositing with these two types of frame buffer only.

\begin{figure*}[!htb]
  \centering
  \includegraphics[width=1.0\linewidth]{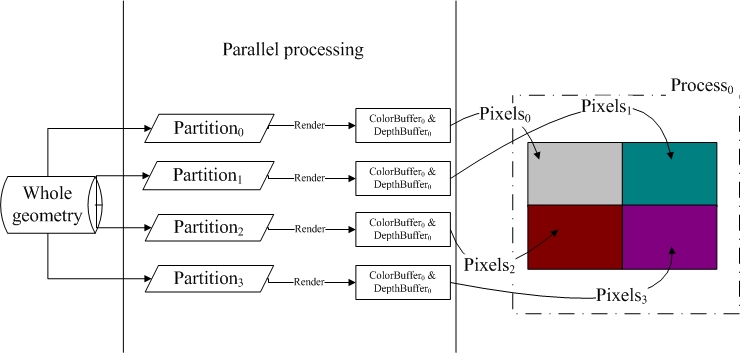}
  \caption{\label{fig:partitioning}
  Illustration of data partitioning and pixel-wise compositing in our parallel visualizations}
\end{figure*}

\begin{figure*}[!htb]
  \centering
  \includegraphics[width=1.0\linewidth]{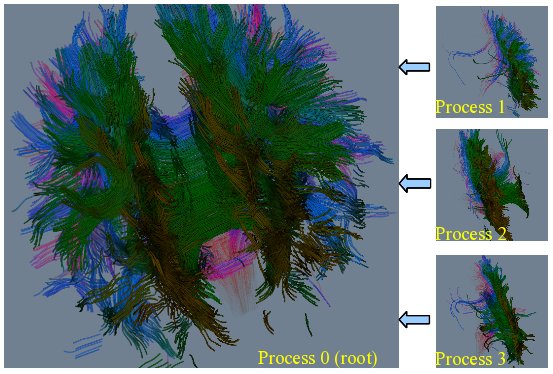}
  \caption{\label{fig:dticompositing}
  An example of the pixel-wise compositing in our parallel visualization scheme. The dense streamtube visualization with
depth mapped to size, color and transparency is parallelized using 4 processes. Process 0 (master) gathers all parallel
renditions from slave processes and composites them together with its own local rendition to produce the complete
rendering.}
\end{figure*}

Procedurally, this compositing process is performed by the following three steps: (1) each process fetches pixels from
the frame buffers in its local process memory space, and (2) all slave (slave, as opposed to the master) processes
send all the buffers one after another to the master process, which do not send its local buffers however. Then, (3) the master
process performs a pair-wise buffer compositing every time it receives the buffer from a slave process until all
slave buffers are composited. Finally, the master process writes the composited depth and color values back to corresponding
local frame buffers as a full image. Figure~\ref{fig:partitioning} illustrates this pixel-wise compositing process, an
example of which is shown in Figure~\ref{fig:dticompositing} using 4 processes to parallelize the rendering task.

In addition, when rendering geometries in parallel in the background, the parallelization should be transparent to
users. So except for special needs for showing slave renditions, no rendering partitions should be visible and
the composited visualization is displayed on the master process only. There are two points to make for the compositing to
be optimized.

First, off-screen rendering is applied to avoid slave renditions. This is not only to meet the
need of slave renderers for invisible renditions but, more importantly, to improve the overall rendering
performance. Second, creation of rendering windows on all slave processes is avoided. Depending on the graphics
platform practically used, a less ideal solution is to hide the rendering windows if the creation of them is required
for correct rendering. Example case is that a window must be created to establish a context for the drawing to take
place. Finally, synchronizing camera parameters across all processes before any process starts to render can simplify
the later process of image compositing. As adopted in our approach, a simple way of synchronization is to broadcast the
key camera parameters (focal point and position) retrieved on the master process to all slave processes.

\subsection{Depth Mappings}
Depth mappings are applied to stylize geometry unit according to its depth information so that a better perception in
the 3D environment can be obtained. Depending on how the depth value is mapped to the value of different visual
variables, a depth mapping is either a linear or non-linear function $\displaystyle f(v) = V(Rank(v_d))$ where
$\displaystyle Rank(x)$ is the order of $x$ in the sorted sequence and $v_d$ is the depth value of a single geometry
unit (vertex will be consistently exemplified in the following text), and function $V$ maps the ranking order sequence
to the range of designated visual variable, $\displaystyle [V_{min},V_{max}]$. In the case of linear mapping, for instance,
\begin{equation}
\displaystyle V(x) = \frac{V_{max} - V_{min}}{x_{max} - x_{min}} (x - x_{min})
\end{equation}

As we consider size ($s$), color ($c$), value ($i$) and transparency ($t$) as the variables mapped, $V(x)$ is a scalar
function. Further, $V(x)$ is unitary function for single mapping and multiple mappings are simply an aggregation of
multiple single mappings. For example, when mapping depth to size, color and transparency at the same time, $\displaystyle
V:x\rightarrow(s,c,t)$ is essentially $\displaystyle V:x\rightarrow(S(x),C(x),T(x))$ where $S,C,T$ are all unitary
mappings.

\subsubsection{Depth Mappings in Parallel Rendering}
\label{sec:DepthMappings}
In the context of geometry rendering, depth mappings are easily performed according to the simple function
evaluations as described above. However, depth mappings need be parallelized as well in order to collaborate with
parallel rendering towards an optimized performance in the context of visualization parallelization. In our parallel
visualization, depth mappings are required to be coherent in the geometry model a whole. Therefore, simply mapping local
geometry on each process independently and then compositing the locally depth-stylized renditions do not work.

For each process, the input of depth mapping is the ranking order of depth values of local geometries and, as a result,
each process will only have the local rank for every vertex in local its geometry partition. However, the global rank of
a vertex in the range of the whole geometry must be retrieved for a coherent global depth mapping. With global ranks of
local vertices, every process can render its local geometry independently yet correctly due to the correct mappings from
the local vertices to the partition of the range of $\displaystyle V(Rank(v_d))$ corresponding to those vertices.

The following figure shows the outline of the integrated parallel sorting and depth mapping algorithm adopted in our
parallel visualization.
\begin{algorithm}[h]
    \caption{integrated parallel depth sorting and mapping}
    \label{sortingmapping}
    \begin{algorithmic}[1]
        \STATE $numProcs\gets$ total of processes
        \STATE $myId\gets$ local process rank
        \STATE $numPts\gets$ total number of vertices in local partition
        \STATE Gather all $numPts$ values into array $allNumPts$
        \STATE $idoset\gets 0$
        \FOR{$i = 0 \to myId-1$}
            \STATE $idoset\gets idoset + allNumPts[i]$
        \ENDFOR
        \FOR{$i=0 \to numPts-1$}
            \STATE $depth[i].vd\gets$ depth value of the $i$th vertex in local geometries calculated from camera parameters
            \STATE $depth[i].id\gets i+idoset$
        \ENDFOR
        \STATE sort $depth$ according to the $vd$ field using $qsort$
        \STATE Sum up all $numPts$ to $totalPts$
        \IF{$myId == 0$}
            \STATE $oset\gets 0$
            \STATE $tdepth[0..numPts-1]\gets depth[0..numPts-1]$
            \FOR{$i=1 \to numProcs-1$}
                \STATE Receive $tdepth[numPts+oset.numPts+oset+allNumPts[i]]$ from process $i$
                \STATE inplace merge $tdepth[0..numPts+oset.numPts+oset+allNumPts[i]]$
                \STATE $oset\gets oset+allNumPts[i]$
            \ENDFOR
            \FOR{$i=0 \to totalPts-1$}
                \STATE $hashIndex[ tdepth[i].id ]\gets i$
            \ENDFOR
            \STATE Broadcast $hashIndex$
        \ELSE
            \STATE Send $depth$ to master process $0$
            \STATE Receive $hashIndex$ from master process $0$
        \ENDIF
        \FOR{$i=0 \to numPts-1$}
            \STATE $Rank_{global}[i]\gets hashIndex[i+idoset]$
        \ENDFOR
    \end{algorithmic}
\end{algorithm}

\subsubsection{Hash Index}
To figure out the global rank of depth value for each local vertex, reducing all locally sorted partitions
together on the master process and then broadcast the resulting depth array that is globally sorted to all other
processes can be an easy solution. As such, the global rank will be retrieved from the depth array received for
evaluating $f(v)$ for each vertex. However, the retrieval would be a $O(N^2)$ search for all $N$ local vertices, which
is prohibitive enough to make an interactive frame update impossible alone according to our tests. For a real-time
global rank retrieval, we create a global hash index for the whole geometry immediately after the depth sort on the master
process is all finished.

\begin{figure}[htb]
  \centering
  \includegraphics[width=1.0\linewidth]{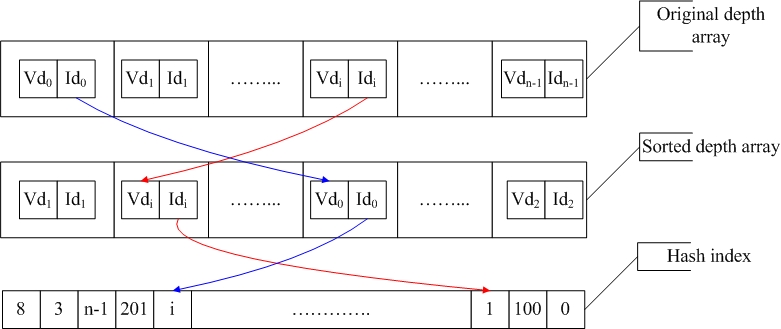}
  \caption{\label{fig:hashindex}
  Hash index for real-time depth mapping in the context of the presented parallel visualization.}
\end{figure}

In both the local and global depth arrays, an index is kept for each depth value at each element and the depth array is
actually a sequence of vector $(d,Id)$ where $d$ is the depth value and $Id$ is the index, which is initialized
with the original global rank of a vertex in the unsorted holistic geometry. As such, wherever a depth array element is
moved after sorting, its original rank, taken as a vertex identifier as well, can be always retrieved immediately. We
use this id to associate the unsorted and sorted depth array through the hash index. Figure~\ref{fig:hashindex} illustrates
the hashing process for depth mapping.

\subsubsection{Mapping Update}
During the interactive exploration of the depth stylized visualization, mappings need be updated whenever the depth
order of geometry along the viewing direction changes, typical because of data rotation. The mapping update is then
reflected through refreshed rendering, which is right the reason why we explore the rendering parallelization as discussed
before. It is reasonable, therefore, to actively trigger the frame update once mapping update has taken place. There are
at least two different mechanisms for the frame update to be timely invoked by mapping updates. First of all, a
polygonal data filter, which is used for the purpose of depth sorting, can be inserted into the demand-driven rendering pipeline
so that rendering update will be triggered when either the input or output of the filter is modified (using VTK is in
this case).

However, besides updating depth mappings, geometry copy between the data filters is required, which has been
proven to heavily drag the frame rate. It is noteworthy here that the geometry is not really updated at all when the
depth mappings change. Another mechanism is to explicitly invoke frame update (redrawing) through interaction handling.
With this approach, only mappings are recomputed while no geometry copying is involved. We employ the latter for a
better performance.

In the interaction-driven mapping update mechanism, we only directly handle user input, such as mouse interaction, that
shuffles the depth order of geometries on the master process. when responding to such user input, the master process
invokes frame update after finishing mapping calculations and then sends a remote method invocation (RMI) message to all
slave processes. In the RMI handler on each process, mapping update is firstly triggered, followed by an active call
to frame update. Apparently, there is a message processing loop on all processes to enable real-time RMI responding.

\section{Implementations}
Our parallel depth stylized visualization is implemented in C/C++ using VTK with parallelism support by MPI. In the
parallel sorting algorithm, $qsort$ routine from the standard C library for local quick sort on each process and generic
in-place merge algorithm in C++ STL library for iterative two-way merge sort on the master process. We have employed the
image compositing functionalities provided by VTK's parallel modules but extended certain classes to tailor their
functions for our customized pipeline components in order to implement the pixel-wise compositing. While off-screen
rendering has been directly supported in VTK, we make use of wrapping windows by Qt widgets to hide rendering windows of all
slave processes.

In addition, our depth sorting filter is extended from VTK's polygonal data depth sort filter and an interactor component
extended from VTK's track-ball camera interactor, which work together to meet our needs for the interaction-driven
mapping update. To explicitly trigger frame update, user-defined RMI messages for this purpose are added and the
callbacks are registered to VTK's multiple process controller component before parallel rendering starts. With these
extended components, the interactor responds to data rotation by broadcasting mapping update RMI message to all slave
processes and then mapping calculations and frame update are invoked in the callback of the RMI message. The
visualization program is simply running as a MPI application and thus the number of processes can be indicated when
launching the MPI runtime. As we discuss in detail in section ~\ref{sec:Results}, an optimal number of processes to be
indicated depends on the actual hardware architecture.

\begin{figure}[htb]
  \centering
  \includegraphics[width=1.0\linewidth]{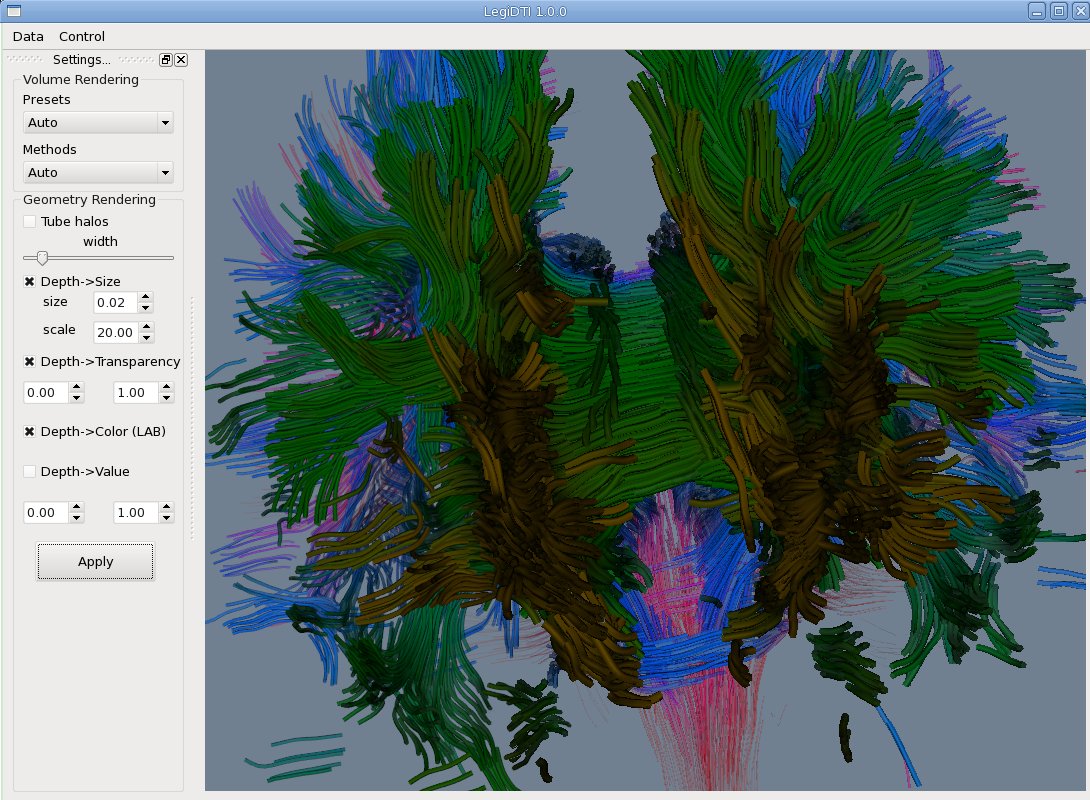}
  \caption{\label{fig:legidti}
  The outlook of the depth-stylized 3D tube visualization parallelized using the proposed method.
}
\end{figure}

Figure~\ref{fig:legidti} shows the outlook of our test application of the presented parallel visualization method.
The GUI framework is created using Qt 4.0 by which all the interaction widgets are set up for the depth stylizing
customization. In order to achieve an optimized performance,
parallel processing is only applied to the rendering widget and all other GUIs are created on the master process only.

As such, GUI interactions have to be explicitly relayed from the master process where they are triggered to all slave
processes so that the slave rendering can reflect the changes in the stylizing configuration as the result of those
interactions. We register another type of RMI message and define a dedicated callback to realize the RMI for updating
slave renderings. RMI messages are easily transmitted by MPI communications.

\section{Results}\label{sec:Results}
\begin{figure*}
  \centering
  \subfigure{
  \includegraphics[width=.46\linewidth]{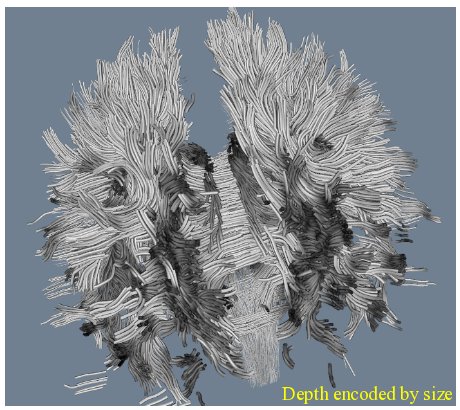}
  }
  \subfigure{
  \includegraphics[width=.46\linewidth]{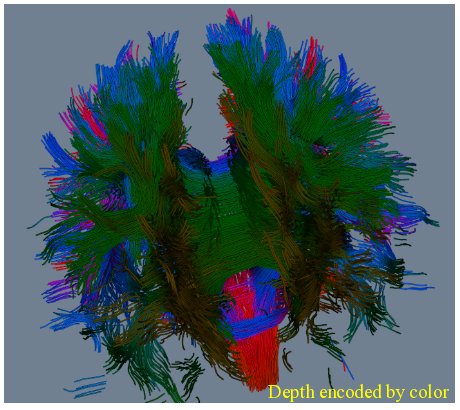}
  }
  \subfigure{
  \includegraphics[width=.46\linewidth]{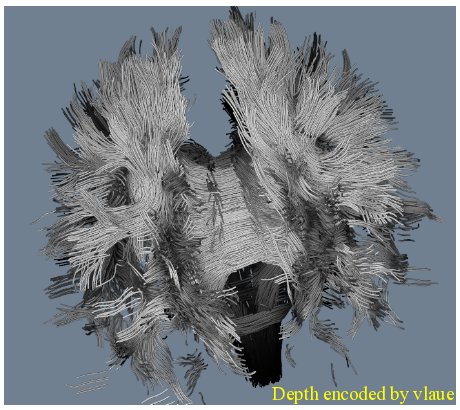}
  }
  \subfigure{
  \includegraphics[width=.46\linewidth]{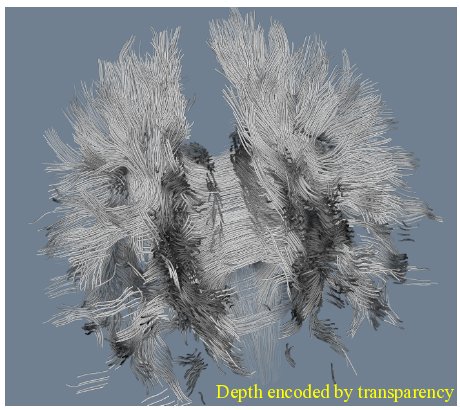}
  }
  \subfigure{
  \includegraphics[width=.46\linewidth]{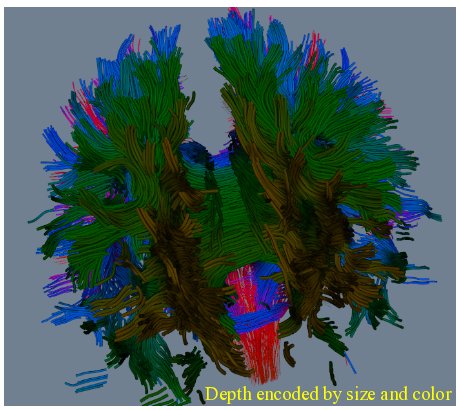}
  }
  \subfigure{
  \includegraphics[width=.46\linewidth]{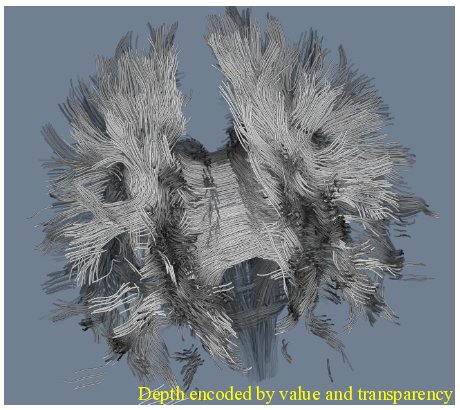}
  }
  \caption{\label{fig:dtidepthmappings}
  Our parallel visualization of DTI streamtubes with single mapping including depth to size (upper left), color (upper right), value (middle left) and transparency (middle right) respectively, and multiple mappings including depth to size and color combined (bottom left) and to value and transparency combined (bottom right). We use these different mappings with typical visual variables to communicate depth information in the 3D visualizations.
  }
  \vspace{-2pt}
\end{figure*}

We have applied the parallel visualization pipeline presented in this paper to interactive depth-stylized visualization
for the purpose of investigating legibility issues in 3D data visualizations. As our current application scenarios, we
create streamtube visualizations of diffusion tensor MRI (DTI) data with single and multiple depth mappings applied in order
to enhance users' depth perception in the 3D visualizations, as shown in Figure~\ref{fig:dtidepthmappings}.

Also, on the basis of above implementations, we evaluate the efficiency of our parallel visualization approach by firstly
measuring the overall rendering performance including depth sorting and MPI communication costs and then comparing our
method to other alternative parallel rendering implementations. Our evaluation is based on the results collected from
many runs of our test application shown before on a Intel(R) Core(TM)2 Quad 2.66GHz processor with 4GB DDR2 memory.
\subsection{Performance Measurements}\label{sec:PerformanceMeasurements}
We measure the proposed parallel visualization method by first comparing visualization performance of the parallel
approach to the sequential one with different scales of geometry. Precisely, for each one of the test data sets, the time spent by
rendering a single frame in the parallel visualization in milliseconds is paired for comparison with that spent by
the same task in sequential one. Here in our application scenarios, we visualize 3D depth stylized streamtubes generated from
diffusion tensor MRI data with different depth mapping schemes applied for the tests.

\begin{figure}[htb]
  \centering
  \includegraphics[width=1.0\linewidth]{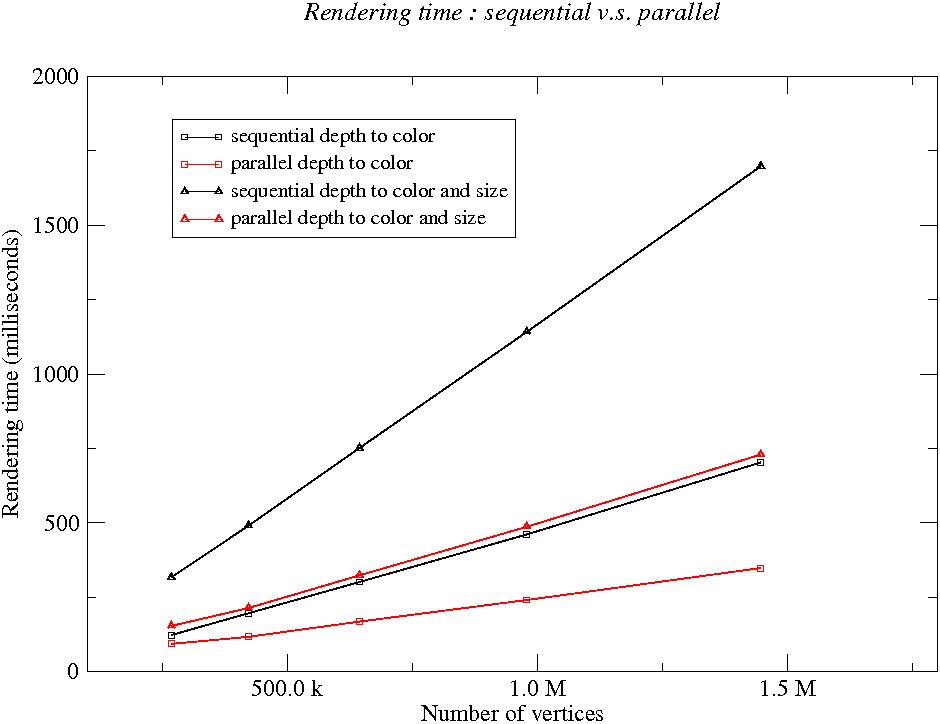}
  \caption{\label{fig:compcost}
  Rendering performance of depth-stylized tube visualization using our parallelizing method compared with
sequential visualization performance. Both single depth mapping and multiple mappings are tested and compared between
the parallel and sequential visualization.}
\end{figure}

As shown in Figure~\ref{fig:compcost}, parallelization enables an interactive rendering performance for our depth-stylized geometry
visualization, which is hard to obtain with sequential approach. Each value of the rendering time measurements is an
average of the total rendering cost over 100 continuous frames. For the parallel rendering, time measured has included
costs of communications among the 4 processes used.

We differentiate only two instances of depth mappings here, depth to color alone and depth to both size and color, because they are
representative of two disparate amount of computations for depth mappings in our tests. For the single mapping from depth to color,
there is only one round of depth sorting besides the rendering task involved. For the multiple mappings from depth to size and color, there are two rounds of depth sorting plus the tube mesh generation besides rendering task. Among the two rounds of depth sorting, one is for depth of line geometry to size mapping before tubes of different radii are generated and the another for
depth of tube geometry to color mapping after tubes are produced from polylines.

To examine how the number of processes used in the parallelization effects the parallel visualization performance,
different values of the number has been tested. As the result in Table~\ref{tab:speedup} presents, performance
increases monotonically along the increase in the number of processes before the number reaches 4, after which the
performance decreases also monotonically when the number continues to grow. That the maximal speedup is achieved at the
number of processes of 4 can be attributed to the fact that the number of hard CPU cores is 4.
\begin{table}
\centering
\begin{tabular}{|c|c|c|c|c|c|c|}
\hline
\multirow{2}{*}{Metrics}	& \multicolumn{6}{c|}{Number of Processes} \\
\hhline{~------}
              & 2	  & 3      & 4      & 5      & 8     & 12        \\
\hline
Time (ms)		  & 409   & 359 & 347 & 401 & 469 & 642  \\
Speedup		  & 1.72  & 1.95   & 2.02   & 1.75   & 1.5   & 1.09     \\
Efficiency	  & 0.86  & 0.65   & 0.51   & 0.35   & 0.19  & 0.09     \\
\hline
\end{tabular}
\caption{The effect of the number of processes employed on the parallel performance gauged by time cost in milliseconds, parallel
speedup and efficiency, tested using our parallel approach with the visualization with depth to color mapping of 9,635
tubes including 1,447,005 vertices.}
\label{tab:speedup}
\end{table}

\subsection{Comparisons}
we further verify the efficiency of our parallelization approach for depth-stylized geometry rendering by comparing the
the overall visualization performance gained by our method with that by other alternative approaches. We implemented the
3D tube visualization with depth-stylizing using both a partially and a fully parallelized rendering. For both
comparisons, we gauge the total rendering time with five different scales of 3D tube geometries stylized by depth-dependent color
and color, similar to the methodology for measuring performance gain of parallel over sequential visualizations described in
section ~\ref{sec:PerformanceMeasurements}. Constantly, 4 processes are used in all the following tests.

\begin{figure}[htb]
  \centering
  \includegraphics[width=1.0\linewidth]{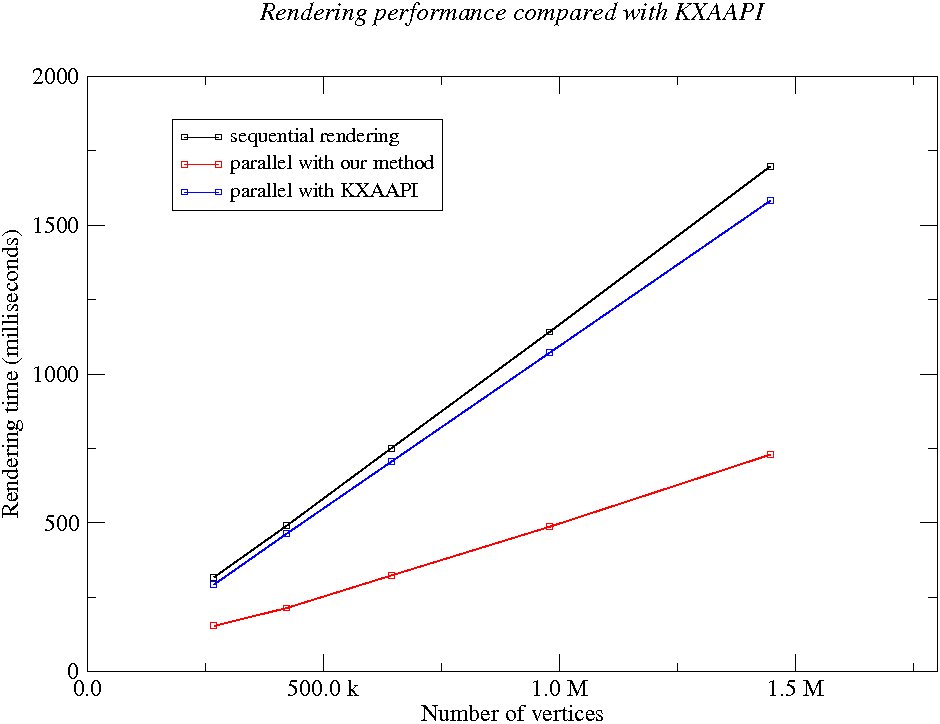}
  \caption{\label{fig:kxaapi}
  Rendering performance of depth-stylized tube visualization using our fully parallelized pipeline compared with that of
the same visualization using KXAAPI for depth sorting in the partially parallel visualization. Performance of sequential
rendering is also included for comparisons.
}
\end{figure}

By comparing to the partially parallel rendering, in which only the depth sorting is parallelized while the overall
rendering pipeline is sequential, we intend to show the advantages of our approach with respect to meshing the sorting
parallelization with the rendering parallelization. We employed the Kernel for Adaptive, Asynchronous Parallel
and Interactive programming (KXAAPI) library ~\cite{KXAAPIUrl} to sort the depth information of the whole geometry on
the sequential visualization pipeline of VTK. As the results, Figure~\ref{fig:kxaapi} shows our approach appears much superior
and outperforms by more than two folds.

\begin{figure}[htb]
  \centering
  \includegraphics[width=1.0\linewidth]{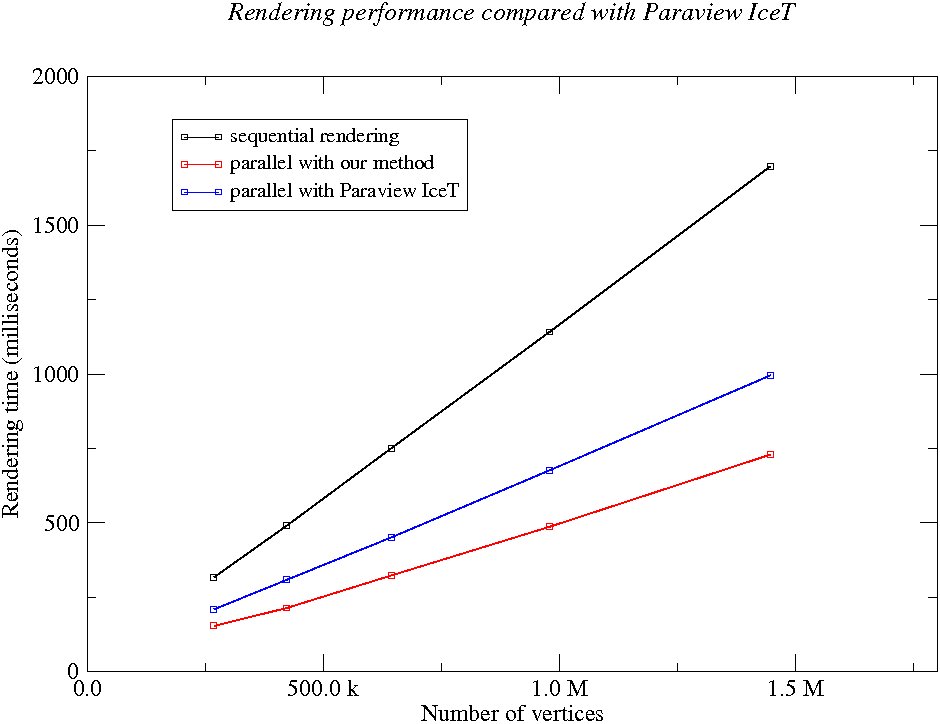}
  \caption{Rendering performance of depth-stylized tube visualization using our parallel approach compared with that of
the same visualization using Paraview IceT, both being fully parallelized pipeline, including sequential performance or comparisons.}
\label{fig:paraviewIceT}
\end{figure}

For an alternative fully parallelized visualization solution to compare, we implemented our depth-stylized tube
visualization using the IceT module in Paraview ~\cite{IceTUrl} with the same measuring method as above. We have
partially ported the IceT module from Paraview source package into VTK for the purpose of our test. As is shown in
Figure~\ref{fig:paraviewIceT}, our parallel visualization solution definitely outperforms the IceT based parallelization
scheme.

\section{Discussion}
In our application scenario, we visualize depth-stylized 3D tubes, which is generated in the run-time by wrapping
polylines that is loaded into our parallel visualization pipeline. Alternatively, tube meshes can be produced prior to
its being loaded to the pipeline then we the computational cost for tube generation will be eliminated. There are two reasons for not doing the off-line tube generation.

First, we need to change the radii of tubes to reflect the depth
changes for the depth to size mapping. Loading line geometries and then generating tubes in the run-time is a more
efficient way for visualizing tubes with depth to tube size mapping than loading tube meshes directly and then transforming
each tube to implement the mapping.

Second, as mentioned in section ~\ref{sec:PerformanceMeasurements}, by means of
online tube generation, we differentiate two types of mappings, single and multiple, in terms of computational costs in
order to demonstrate that a parallel visualization approach like ours will achieve a even superior overall visualization
performance if they are more computational steps, such as geometry processing like wrapping poly-lines to produce tube
meshes, involved within the whole geometry rendering job.

As a matter of fact, it is obviously shown in the performance measurement results that visualization accelerations are much greater with multiple mappings from depth to size and color than with single mapping from depth to color when the geometry scale is increasing. This is due to that more computations are to be parallelized as well and thus the overall performance gain by parallelization increases compared to sequential visualization. According to this analysis, it is reasonable to scale our parallel visualization method to a more complex visualization context where more compute-intensive steps associated with the rendering task must be involved. Results from our tests before have initially show this type of scalability of our proposed approach.

In addition, although we currently use only a single four-core processor to test our parallelization scheme, it is
reasonable to predict that the performance speedup shown in Table ~\ref{tab:speedup} will continue to grow if the number of CPU
cores further increases. Also, because of the performance scalability of the underlying MPI facilities to processor architecture,
application of our method to a multiple processor machine can gain even greater visualization accelerations.

\section{Conclusions}
We presented a parallel visualization method that enables real-time floating-point computations involved in depth
mappings for more legible 3D data visualizations via enhanced depth perception, and therefore helps achieve
interactive frame rate in the depth-stylized visualization of large 3D geometries. The method presented is built upon
the MPI paradigm within VTK with necessary extension adopted for vertex depth reordering optimizations. Our approach
has been tested with 3D dense tubes containing millions of vertices with multiple mappings of depth information applied
and the interactive frame rate achieved has shown that our method is efficient for addressing performance issues inherent
in the visualization scenarios exemplified in the depth-stylized visualizations. Nevertheless, our method can be easily
extended to parallelize visualizations of other large-scale geometry data where intensive computations are required in order to obtain interactive rendering speed.

We have demonstrated the superior efficiency of our approach as a CPU-based parallel visualization framework by comparing the real-time rendering performance of the method presented with that of both sequential method and other parallelization approaches such as XKAAPI and Paraview Icet. As the results show, the proposed framework can provide an efficient alternative to parallel visualization solutions relying on high-end hardware such as GPUs for interactively visualizing large-scale 3D geometry models like stylized dense tubes when the high-end hardware is not readily available. 

\bibliographystyle{abbrv}
\bibliography{paper_arXiv}

\end{document}